\documentclass[aps, prl, reprint, letterpaper, longbibliography, superscriptaddress]{revtex4-1}

\RequirePackage{graphicx}
\usepackage{xfrac}
\usepackage[cmex10]{amsmath}
\DeclareMathOperator\e{e}
\DeclareMathOperator\W{W}

\begin{document}                  



\title{Dead-time Correction for Spectroscopic Photon Counting Pixel Detectors}

\author{Gabriel Blaj}
\affiliation{SLAC National Accelerator Laboratory, 2575 Sand Hill Road, Menlo Park, CA 94025, USA}
\email{blaj@slac.stanford.edu}





\begin{abstract}
Modern photon counting pixel detectors enabled a revolution in applications at synchrotron light sources and beyond in the last decade. One of the limitations of current detectors is reduced counting linearity or even paralysis at high counting rates, due to dead-time which results in photon pile-up. Existing dead-time and pile-up models fail to reproduce the complexity of dead-time effects on photon counting, resulting in empirical calibrations for particular detectors at best, imprecise linearization methods, or no linearization. This problem will increase in the future as many synchrotron light sources plan significant brilliance upgrades and free-electron lasers plan moving to a quasi-continuous operation mode. We present here the first models that use the actual behavior of the analog pre-amplifiers in spectroscopic photon counting pixel detectors with constant current discharge (e.g., Medipix family of detectors) to deduce more accurate analytical models and optimal linearization methods. In particular, for detectors with at least two counters per pixel, we completely eliminate the need of calibration, or previous knowledge of the detector and beam parameters (dead-time, integration time, large sets of synchrotron filling patterns). This is summarized in several models with increasing complexity and accuracy. Finally, we present a general empirical approach applicable to any particular cases where the analytical approach is not sufficiently precise.

\end{abstract}

\date{\today}

\maketitle 

     
\section{Introduction}

Modern photon science was made possible by the advent of hybrid pixel detectors \cite{heijne1988silicon,anghinolfi19921006}, leveraging developments in commercial ASIC design while separating it from sensor design. This enabled almost two decades ago the first highly segmented photon counting pixel detectors, with 64~K photon counting pixels per ASIC \cite{llopart2001medipix2}. Entire research fields would be unrecognizable without the current wide availability of photon counting detectors: protein crystallography \cite{broennimann2006protein}, industrial X-ray diffraction \cite{devries2007medipix}, electron microscopy \cite{sikharulidze2011low}, medical imaging \cite{taguchi2013vision}, computed tomography \cite{pollmann2010multidimensional}.

One of the limitations of current photon counting pixel detectors is the loss of linearity at high counting rates, followed by saturation and paralysis at even higher photon rates \cite{walko2008empirical}. This problem is set to increase in the near future, as many synchrotron facilities are planning substantial upgrades in brilliance \cite{chenevier2018esrf}, and free electron laser facilities consider operating in a quasi-continuous wave mode in the future \cite{brinkmann2014prospects,marcus2017lcls}.

Ultrafast integrating pixel detectors currently under development will be able to operate at 100~kHz and count linearly at extremely high photon fluxes towards $10^8$~photons/pixel/second \cite{blaj2015xray,blaj2015future,blaj2018performance}, while also providing full spectral information on a photon by photon basis (at low photon occupancies) \cite{blaj2016xray}. However, until these detectors become available and practical for applications outside the demanding area of free-electron laser applications \cite{graafsma2009requirements}, it is useful to improve the performance of currently available photon counting \cite{ballabriga2016review} or time-over-threshold (TOT) detectors \cite{llopart2007timepix,poikela2014timepix3,johnson2014eiger}.

One approach to linearize the saturation behavior of photon counting detectors is to model the dead-time in detectors with a single counter and calculate a transformation function for linearization. Current models describing the dead-time and counting behavior of photon counting pixel detectors are incomplete, relying on characterizing the response of a single counter in a particular detector as a function of experimental conditions, e.g., Mythen \cite{bergamaschi2011time}, Pilatus \cite{trueb2012improved}, Eiger \cite{johnson2014eiger}; assuming a fixed dead-time \cite{bateman2000effect,knoll2010radiation,taguchi2011modeling}; and discussing the effect of photon energy and threshold on counting with an exponential decay pulse shape for the analog signal \cite{bergamaschi2011time}. Note, however, that these detectors use an exponential decay pulse shaping \cite{bronnimann2014hybrid} and have only one counter \cite{ballabriga2016review}, unlike spectroscopic photon counting detectors.

Several simplified analytical models are discussed in \cite{walko2008empirical}: non-extended dead-time, extended dead-time and ``isolated'' dead-time applicable to, e.g., scintillators and avalanche photo diodes, thus progressing towards explaining the count rate of single photon counting detectors beyond the previous methods; however, the model only takes into account fixed, extendable dead-time. The inadequacies of the current dead-time models and the limited performance in photon counting linearization have been noted in the literature \cite{sobott2013success}.

Other approaches to linearize the saturation behavior of photon counting detectors include (1)~using two thresholds and summing their counting rates, which yields a modest increase in linearity \cite{schmitt2015single}; (2)~designing complex detectors combining photon counting with TOT information \cite{bergamaschi2011time,schmitt2015single}, with performance limited by the precision of TOT measurements \cite{schmitt2015single}; (3)~design detectors with two different shaping times, then model and correct their behavior \cite{abbene2015high}; or (4)~using pixels with two thresholds and counters, and calculating the sum of the lower counter with the multiplication by a fixed factor with the higher counter \cite{kappler2010research,kappler2012first,kraft2012experimental}, similar to (1); note however that the fixed factor is only valid at a particular photon rate.

We present here a statistical approach to modeling dead-time and counting nonlinearities for typical spectroscopic detectors (i.e., with multiple energy thresholds) and a constant current discharge of the feedback capacitor \cite{krummenacher1991pixel,llopart2001medipix2,llopart2007timepix,ballabriga2013medipix3rx} and resulting pile-up counter models. These models reproduce the characteristics of counter rates as a function of photon rates. We show that using two counters at different thresholds results in a doubling of maximum counting rates compared to detectors with a single counter and yields 1 to 2 orders of magnitude increase in linear counting rates for quasi-continuous light sources, while eliminating the need for calibration or knowledge of particular detector parameters (frame shutter time, dead-time) in detectors with two (or more) thresholds. Finally, we discuss the effect of typical synchrotron fill patterns on the counting rate. This is relevant for the entire community of photon counting detector research, applications, and their users \cite{ballabriga2016review}.

\section{Photon Counting Pixel Detectors} \label{sec.2}

\subsection{Analog Signal} \label{sec.2.1}

A sensor (typically a reverse biased p-n junction) detects a photon which is converted in electron-hole pairs. Depending on the sensor, holes (usually, for n-type silicon sensors) or electrons drift toward the pixel flip-chip bonding pads \cite{blaj2017analytical}. The charge is collected, amplified, and converted to a voltage in the charge preamplifier of each pixel \cite{llopart2001medipix2}. For simplicity, we will call the analog output voltage of the charge preamplifier the ``signal''.

\begin{figure}[t]
    \includegraphics[width=\columnwidth]{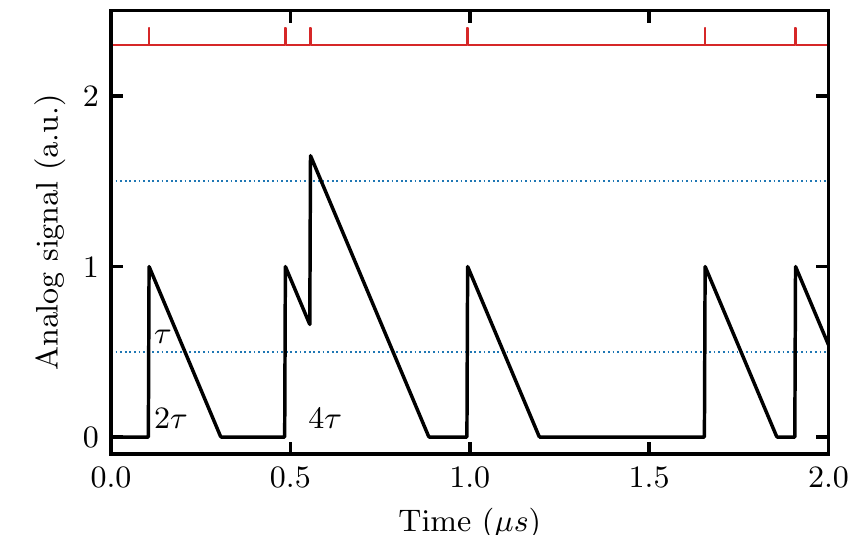}
    \caption{Example of typical signal presented to the threshold comparator, resulting from a relatively short shaping time and a constant discharge current (black line), corresponding to discrete photon arrival times (red line). The two comparator thresholds ($\sfrac{1}{2}$ and $\sfrac{3}{2}$) are indicated by the blue dotted lines. Typically, individual photons only trigger the lower threshold (most photons shown), however, occasionally the time difference between two photons is short enough to lead to pile-up and triggering of the higher threshold (near $t=0.6~\mu s$). The example shown here corresponds to a normalized photon rate $2\lambda\tau=0.2$, where $2\tau$ is the decay time of the analog signal from $1$ to $0$.}
    \label{fig1}
\end{figure}

Typically, the preamplifier is implemented with a constant current discharge of the feedback capacitor \cite{krummenacher1991pixel,llopart2001medipix2}, resulting in a approximately triangular pulse shape as depicted in Fig.~\ref{fig1}. This pulse is then compared with one or more thresholds \cite{ballabriga2013medipix3rx}.

\subsection{Photon Counting} \label{sec.2.2}

One or more threshold comparators are connected to corresponding counters; when the signal exceeds one of the thresholds, the corresponding counter is incremented by 1 \cite{llopart2001medipix2}. The ``photon counting'' approach thus discards small signals and corresponding noise (along with detailed spectral information of individual photons), providing a compact, lossy summary of data \cite{tlustos2005performance,blaj2016xray,michalowska2017detection}; the noise is still present (over-imposed on photon signals), leading to a relatively limited spectroscopic resolution of photon counting detectors \cite{tlustos2005performance}.

The thresholds must be set judiciously to minimize the effect of charge sharing along the edges between two pixels; this corresponds to a setting of $\sfrac{1}{2}$ of expected photon signal amplitude for detectors with a single threshold, or settings of $\sfrac{1}{2}$ and $\sfrac{3}{2}$ of expected photon signal amplitude for detectors with 2 thresholds (see dotted lines, Fig.~\ref{fig1}). When each of these thresholds is exceeded, the corresponding counters $C_0$ and $C_1$ are incremented by $1$. Note that when set at the default value of $\sfrac{3}{2}$, $C_2$ can only be triggered by pile-up, i.e., two photons arriving within a short time $\tau$.

In most photon counting pixel detectors this results in reduced quantum efficiency at the corners of 4 pixels, unless a complex ``charge summing'' mode is implemented \cite{ballabriga2013medipix3rx}. More than 2 thresholds are available in some detectors, e.g., in Medipix3RX, up to 8 \cite{ballabriga2013medipix3rx}; see review in \cite{ballabriga2016review}.

\subsection{Photon Pile-up} \label{sec.2.3}

With typical X-ray sources of constant intensity, photon detection is a Poisson process; the time interval between successive photon arrival times $t$ follow an exponential distribution $\sim \lambda\e^{-\lambda t}$, where $\lambda$ is the average photon rate per pixel.

Single photon counts are typically obtained as $C_0-C_1$ (i.e., by subtracting photon pile-up counts $C_1$ from the total number of detected events $C_0$). Note that multiple comparators threshold the same signal, thus the noise corresponding to the counting statistics is limited to $1/\sqrt{C_0-C_1}$. Photon counting detectors with a single threshold and counter could perform a two threshold measurement sequentially, however, $C_0$ and $C_1$ would be statistically independent, leading to larger errors (in the order of $\frac{\sqrt{C_0+C_1}}{C_0-C_1}$).

Photon pile-up can result either in peak pile-up (when photons arrive at very short intervals), triggering comparators corresponding to the sum of their energies; or they can result in spectral distortion when a second photon arrives on the tail of the pulse of a previous photon \cite{ballabriga2016review}. In either case, the pile-up can be modeled, resulting in methods to linearize the response of the counters.

In applications with a complex spectrum, the photon pile-up can be difficult to recover \cite{blaj2017optimal}, especially with the lossy information collected by photon counting detectors. Spectroscopic detectors with two or more thresholds and counters could be used to recover some of the lost information.

\begin{figure}[t]
    \includegraphics[width=\columnwidth]{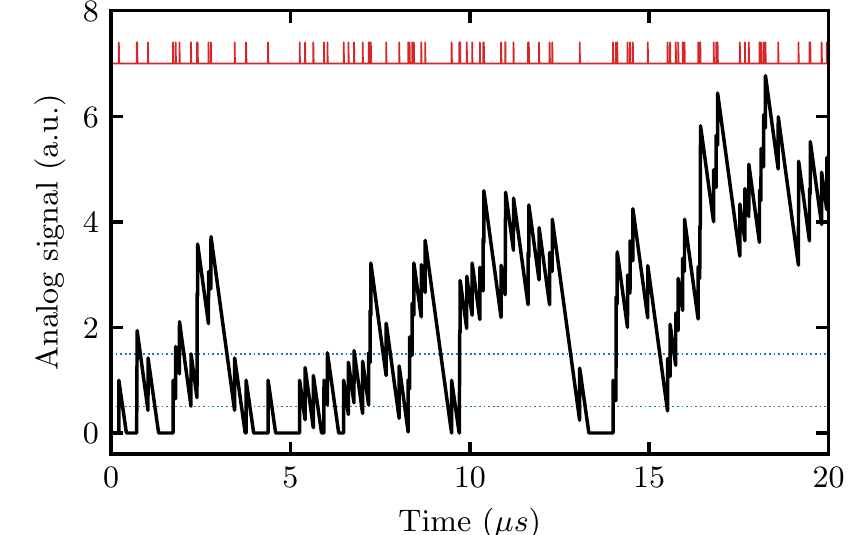}
    \caption{Example of significant pile-up corresponding to a photon rate of $2\lambda\tau=1$, where $2\tau$ is the signal decay time of the analog signal from $1$ to $0$. Note the resemblance to a random walk, resulting in an apparent baseline shift and the complete breakdown of counting in large stretches of time.}
    \label{fig2}
\end{figure}

However, many of the applications with the highest dynamic ranges use (quasi-)monochromatic radiation, e.g., protein crystallography \cite{broennimann2006protein}, X-ray reflectometry \cite{devries2007medipix}, X-ray computed tomography \cite{pollmann2010multidimensional}, X-ray imaging \cite{procz2011flatfield}, and wavelength dispersive spectrometry \cite{blaj2018hammerhead}; this simplifies considerably the task of estimating the effect of photon pile-up on the counters and linearizing the pixel response.

\subsection{Simulation of Pile-up and Effects on Counters} \label{sec.2.4}

To test the analytical correction methods in section~3, we implemented a Monte Carlo simulator for estimating and modeling the effect of pile-up on the counters of photon counting pixel detectors, using several assumptions: (1)~monochromatic radiation, (2)~constant discharge current of the feedback capacitor, resulting in triangular pulse shape with abrupt onset; (3)~fast shaping time compared to discharge time, resulting in a triangular pulse shape; (4)~two thresholds set at $\sfrac{1}{2}$ and $\sfrac{3}{2}$ of the pulse height for the photons, corresponding to counters $C_0$ and $C_1$, respectively.

\begin{figure}[t]
    \includegraphics[width=\columnwidth]{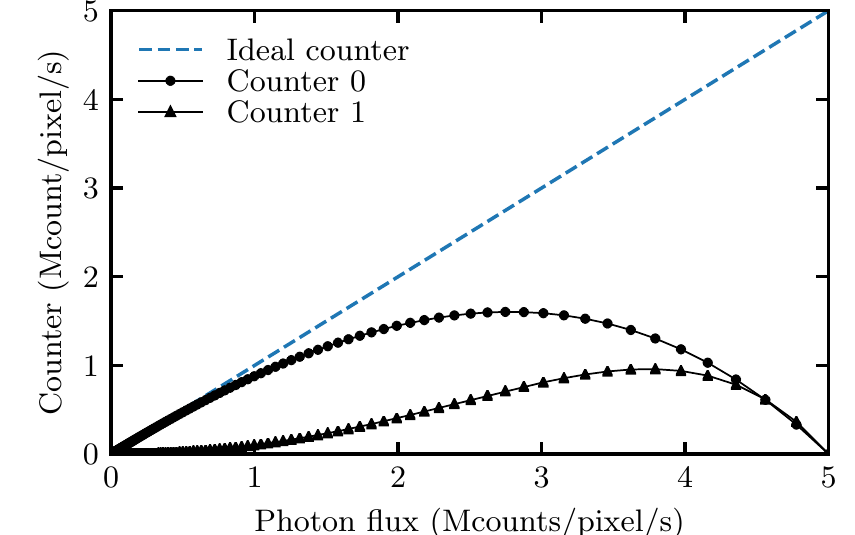}
    \caption{Simulated counter rate as a function of photon flux with $T$=10~s, $\tau$=100~ns. Both counters $C_0$ and $C_1$ (corresponding to thresholds of $\sfrac{1}{2}$ and $\sfrac{3}{2}$, respectively) have a complex dependency on the actual photon rate $\lambda$ and deviate considerably from the ideal response (blue dashed line). When $2\lambda\tau$ exceeds $1$ (corresponding to 5~Mcounts/pixel/s), the circuit is driven into saturation, resulting in counting paralysis. Note that using a single counter results in two different photon fluxes corresponding to a single counter value; using a second counter solves the confusion and effectively doubles the useful range.}
    \label{fig3}
\end{figure}

We simulated frame shutter times of 10~s (for improved statistics), with a time step of 10~ns and a dead-time of 100~ns (i.e., time $\tau$ for the signal to return from 1 to $\sfrac{1}{2}$ and re-enable counting in $C_0$, or half of the time $2 \tau$ for the signal to return to the baseline). Note that the actual dead-time $\tau$ has an approximately linear dependence on photon energy. We simulated a large number of photon rates $\lambda$, varying from $5$~$10^3$ to $5$~$10^6$ photons/pixel/second. The results are shown in Fig.~\ref{fig3}, with the saturation effect clearly visible for both counters. Note the similarities and differences in the shape of counter $C_0$ with experimental measurements in Fig.~6 of \cite{sobott2013success} and in Fig.~5 of \cite{kraft2012experimental}. The similarities are due to photon pile-up while the differences are due to the constant current discharge in Fig.~\ref{fig3}, compared to the exponential discharge.

Note that using a single counter results in two different photon fluxes corresponding to a single counter value; using a second counter solves the confusion and effectively doubles the useful range.

While some detectors might use two thresholds to increment a single counter using counter logic, e.g., \cite{llopart2001medipix2}, their behavior is complex and is beyond the scope of this article. These detectors can be typically turned in single threshold counters by disabling the higher threshold \cite{tlustos2005performance}.

\subsection{Threshold Equalization and Gain Calibration} \label{sec.2.5}

Variations in individual pixels result in a range of different offsets of the signal ``zero'' \cite{ballabriga2016review}, with the offsets typically following a gaussian distribution. The spread can be comparable with the photon energies to be measured. As it cannot be corrected in post-processing (like in charge integrating detectors) and it would severely limit the counting performance, the typical practical solution is to include a number of ``trimming bits'' in each pixel to align the response of individual pixels in the matrix \cite{llopart2001medipix2}. Often this calibration is performed with the noise edge, resulting in an equalization where the pixels end up with, e.g., the $5\sigma$ point in the noise distribution aligned; while this is not optimal, it is easy to perform.

Another effect is pixel gain dispersion \cite{ballabriga2016review} (typically within a few percent); while at low energies its effects are limited (especially with charge summing detectors that minimize the impact of charge sharing), at higher energies and with smaller pixels the effects become more pronounced. In imaging applications, this is often calibrated and corrected by flat field measurements (mixing the actual flat field with the pixel gain and charge sharing).

However, for individual energies of monochromatic beams, the threshold equalization can be adapted to mitigate both pixel offset and gain simultaneously. Ideally, the user should only provide the beam energy, and the corresponding equalization mask would be calculated using previously calibrated dark and gain maps. This could be achieved by carefully calibrating the pixel thresholds with different monochromatic energies or element absorption edges \cite{procz2009optimization,kraft2009characterization}, covering the entire applicable spectrum, and calculating the optimal pixel trim mask from the expected beam energy.

Alternatively, a photon counting detector with both hardware offset and gain equalization has been developed recently \cite{grybos201632k}.

\begin{table}[t]
\caption{Glossary}
\begin{tabular}{r l}
$N$   & Number of photons\footnotemark[1]\\
$E_0$ & Photon energy\\
$0$,$\sfrac{1}{2}$,$1$,$\sfrac{3}{2}$ & Analog signal (normalized to photons)\\ 
$C_0$ & Counter 0 (with signal threshold at $\sfrac{1}{2}$ photon)\\
$C_1$ & Counter 1 (with signal threshold at $\sfrac{3}{2}$ photon)\\
$r$   & Counter ratio $C_1/C_0$\\
$\tau$& Dead-time (proportional to photon energy $E_0$)\\
$2\tau$& Time for signal to return from $1$ to baseline ($0$)\\
$\lambda$ & Photon rate (photons/pixel/s)\\
$\Lambda$ & Photon rate (photons/pixel/pulse)\\
$2\lambda\tau$ & Normalized photon rate $\in [0,1]$\footnotemark[2]\\
$T$   & Frame shutter time\\
$g$   & Relative pixel gain (typically close to $1$)\\
$\delta t$ & Gap between two buckets in regular fill pattern\\
\end{tabular}
\footnotetext[1]{For simplicity, we use the same notation $N$ for the number of photons, regardless of whether it refers to the number of simulated photons, detected photons, expected number, or linearized response; the particular meaning in each equation should be clear from the context.}
\footnotetext[2]{While the normalized photon rate can exceed $1$, it leads to saturation and complete counter paralysis.}
\end{table}

\section{Dead-time Models and Counting Linearization} \label{sec.3}

In the ideal case, we use monochromatic radiation with photon energy $E_0$ and all pixels have the same gain. For simplicity, we assume that the photon energy $E_0$ corresponds to an analog signal of $1$ and the analog signal decay time from $1$ to $0$ is $2\tau$. The detection rate is $\lambda$ photons/pixel/s, and the integration time is $T$ seconds. The expected number of photons per pixel is $N=\lambda T$ (i.e., the weighted average of the corresponding Poisson distribution).

For low photon rates compared to the dead-time, $\lambda \ll 1/\tau$, the individual photon signals will be collected separately; with the usual thresholds at $\sfrac{1}{2}$ and $\sfrac{3}{2}$, counter $C_0$ will contain the number of independently detected photons and counter $C_1$ will contain the number of occasional pile-up events when a photon arrives within time $\tau$ of another.

Counter $C_0$ is incremented each time the signal passes threshold 0 of $\sfrac{1}{2}$. This requires a starting signal under the threshold with at least one photon driving the signal over the $\sfrac{1}{2}$ threshold.

Counter $C_1$ is similarly incremented when the signal exceeds threshold $\sfrac{3}{2}$, i.e., the signal is between $\sfrac{1}{2}$ and $\sfrac{3}{2}$ and one or more photons drive the signal over the $\sfrac{3}{2}$ threshold.

We did not implement a signal saturation limit in this simulation, as it (1)~would depend on detector and gain mode and (2)~would have limited effects at low photon energies (e.g., 8~keV). However, signal saturation enforces an upper limit to the random walk effect shown in Fig.~\ref{fig2}, resulting in more frequent returns to the base line and effectively increasing the accuracy of the models presented here.

\subsection{Typical approach} \label{sec.3.1}

In the typical approach \cite{walko2008empirical}, a fixed paralyzable time $\tau$ is assumed. Photons can arrive within the paralyzable time, leading to pile-up:
\begin{equation} \label{eq1}
    C_1=\lambda T \int_{0}^{\tau}\lambda \e^{-\lambda t} dt = \lambda T (1-\e^{-\lambda \tau})
\end{equation}
or they can arrive after the fixed time $\tau$ yielding:
\begin{equation} \label{eq2}
    C_0=\lambda T \int_{\tau}^{\infty}\lambda \e^{-\lambda t} dt = \lambda T \e^{-\lambda \tau}
\end{equation}
resulting in a relatively simple correction formula \cite{schmitt2015single}:
\begin{equation} \label{eq3}
    N=\lambda T = C_0+C_1
\end{equation}

\begin{figure}[t]
    \includegraphics[width=\columnwidth]{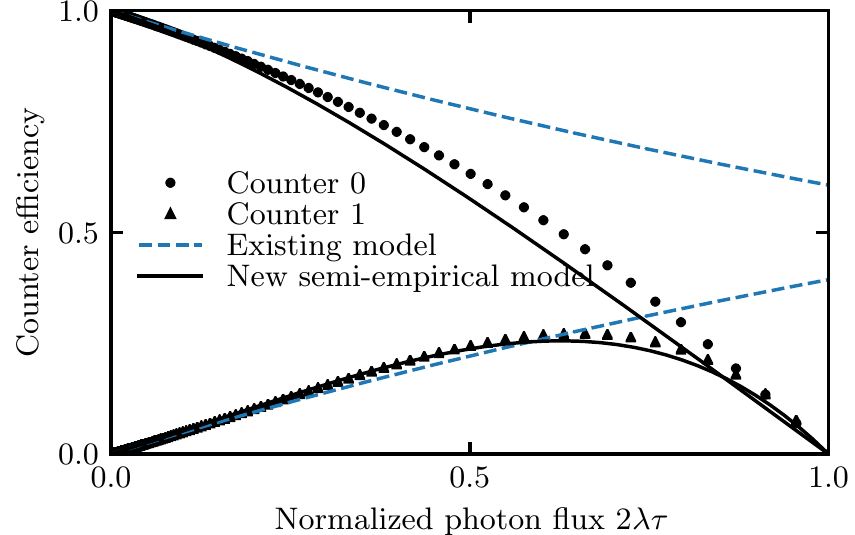}
    \caption{Counter efficiency (i.e., the ratio between counting rates and photon rates) as a function of the normalized photon rate $2\lambda \tau$; an ideal counter would have an efficiency of $1$ over the entire range. The usual models (Eq.~\ref{eq1} and \ref{eq2}), depicted with dashed lines, fails to describe the counter behavior of the two counters at higher photon rates. An improved approximation (Eq.~\ref{eq20} and \ref{eq21}), indicated by the solid black lines, yields much better results and forms the basis of the semi-empirical model described in section~3.4. Systematic deviations are expected, due to the apparent baseline shift at high photon rates.}
    \label{fig4}
\end{figure}

In Fig.~\ref{fig4} we compare the equations with the actual counter behaviors. Note that both Eq.~\ref{eq1} and Eq.~\ref{eq2} are oversimplifications, failing to match the behaviors of $C_0$ and $C_1$ for higher photon rates ($2\lambda\tau$ approaching $\frac{1}{2}$).

\subsection{Novel Simple Model} \label{sec.3.2}

Assuming a photon arriving at $t=0$ (and ignoring the effect of previously arrived photons), the signal will be higher than $\sfrac{1}{2}$ for $t \in [0,\tau)$ and lower than $\sfrac{1}{2}$ for $t \in (\tau,\infty)$. Consequently, counter $C_0$ can be incremented when a second photon arrives and the signal is smaller than $\sfrac{1}{2}$, i.e., at a time $t>\tau$:
\begin{equation} \label{eq4}
    C_0=\lambda T \int_{\tau}^{\infty}\lambda \e^{-\lambda t}dt=\lambda T \e^{-\lambda\tau}
\end{equation}

Incrementing $C_1$ requires exceeding threshold $\sfrac{3}{2}$ which requires at least a third photon arriving within a time $\tau$ of the second photon:
\begin{equation}  \label{eq5}
    C_1=C_0 \int_{0}^{\tau}\lambda \e^{-\lambda t}dt= \lambda T \e^{-\lambda\tau} (1-\e^{-\lambda\tau})
\end{equation}

The ratio of the two is
\begin{equation} \label{eq6}
    r=\frac{C_1}{C_0}=1-\e^{-\lambda\tau}
\end{equation}
with the corresponding number of photons:
\begin{equation} \label{eq7}
    N=\lambda T = \frac{C_0}{\e^{-\lambda\tau}}=\frac{C_0^2}{C_0-C_1}=C_0\frac{1}{1-r}
\end{equation}
This works directly with pixels equalized at energy $E_0$. In case of a different equalization (e.g., with noise, or at a different energy), the (relatively small) variation in gain might play a role, which is discussed in subsection~\ref{sec.3.3}. In this formulation, parameters $\tau$ and $T$ drop gracefully out of the equations.

\subsection{Novel Simple Model with Relative Pixel Gain} \label{sec.3.3}

Typically the thresholds are set globally per pixel; the ideal values for threshold 0 and threshold 1 are $\sfrac{1}{2}$ and $\sfrac{3}{2}$, respectively (see discussion in section~\ref{sec.2.2}). With or without an optimal threshold equalization (see section~\ref{sec.2.5}), small residual differences in relative pixel gains could be present; these can be calibrated with a ``gain map'' \cite{blaj2017robust}. Similarly, a small variance across pixels of the discharge current might be present (Ballabriga 2018, personal communication); the effects of this variation in discharge slope are matematically equivalent to an additional small variation in the pixel gain map.

In this subsection, we investigate the effects of gain variation on the simple model above.

For individual pixel gain $g$ (with $g \to 1$, typically within a few percent), the equations above become:
\begin{equation} \label{eq8}
    C_0=\lambda T \int_{(g-\sfrac{1}{2})2\tau}^{\infty}\lambda \e^{-\lambda t} dt = \e^{-(g-\sfrac{1}{2})2\lambda\tau}
\end{equation}
and
\begin{equation} \label{eq9}
    C_1=C_0 \int_{0}^{(2 g-\sfrac{3}{2})2\tau}\lambda \e^{-\lambda t} dt = C_0 (1-\e^{-(2 g-\sfrac{3}{2})2\lambda\tau})
\end{equation}
with the ratio between the two:
\begin{equation} \label{eq10}
    r=\frac{C_1}{C_0}=1-\e^{-(4 g-3)\lambda\tau}
\end{equation}
resulting in:
\begin{equation} \label{eq11}
    \e^{-(4 g-3)\lambda\tau}=1-r
\end{equation}
by taking the natural logarithm and multiplying by $\frac{2 g-1}{4g-3}$ we obtain:
\begin{equation} \label{eq12}
    -(2 g-1)\lambda\tau=\frac{4 g-3}{2 g-1}\ln{(1-r)}
\end{equation}
which can be rewritten as:
\begin{equation} \label{eq13}
    \e^{-(2 g-1)\lambda\tau}=(1-r)^\frac{4 g-3}{2 g-1}
\end{equation}
Substituting Eq.~\ref{eq13} in Eq.~\ref{eq8} finally yields the linearization:
\begin{equation} \label{eq14}
    N=\frac{C_0}{(1-r)^\frac{4 g-3}{2 g-1}}
\end{equation}
which is the general form of Eq.~\ref{eq7}. Note that Eq.~\ref{eq14} can't be used directly for, e.g., 3 photon pile-up; Eq.~\ref{eq9} would have to be rewritten for the pile-up of 3 photons.

The relative gain $g$ has a relatively small influence on the counting rate correction:
\begin{equation} \label{eq15}
    \lim_{g\to 1}N=\frac{C_0^2}{C_0-C_1}\Bigg[1-2(g-1)\ln\Bigg(1-\frac{C_1}{C_0}\Bigg)\Bigg]
\end{equation}
obtained by Taylor expansion of Eq~\ref{eq14} and keeping the first two terms; the second term is relatively small compared to $1$ because $\lim_{g \to 1} (g-1)$ is small.

\subsection{Novel Semi-Empirical Model} \label{sec.3.4}

The ``simple'' model above performs well at low photon rates $2\lambda\tau \approx \frac{1}{2}$, however, it ignores the stochastic ``baseline shift'' (similar to a random walk) shown in Fig.~\ref{fig2}. We introduce here a simplified approach to deal with this effect.

Due to the triangular pulse shape and linear behavior of the signal, the fraction of time spent above the noise floor is $2\lambda\tau$, thus the fraction of time spent in the noise floor is:
\begin{equation} \label{eq16}
    F_{0}(2\lambda \tau)=1-2\lambda \tau
\end{equation}

Pulses corresponding to exactly $n$ photons have a duration of $2 n \tau$; the probability of detecting exactly $n$ photons in that time interval can be calculated by taking into account the Poisson distribution of $n$ photons in $2 n \tau$ time, modulated by the probability to be in the ``ground state'' ($1-2\lambda\tau$), resulting in a fraction:
\begin{equation} \label{eq17}
    F_{n}(2\lambda \tau)=(1-2\lambda \tau)\frac{(2 n \lambda \tau)^n \e^{-2 n \lambda \tau}}{n!}
\end{equation}

Consequently, the probability of the signal to be under $\sfrac{1}{2}$ is approximately:
\begin{equation} \label{eq18}
    p(s<\sfrac{1}{2})=F_{0}(2\lambda \tau)+\frac{F_{1}(2\lambda \tau)}{2}+...\approx(1-2\lambda \tau)(1+\lambda \tau \e^{-2\lambda \tau})
\end{equation}
(keeping only the first two terms for simplicity). The factor 2 corresponds to the fraction of time with signal $< \sfrac{1}{2}$, i.e., $\tau$ during pulse duration $2 \tau$; using additional terms improves the model accuracy (not shown).

Multiplying with the probability of a signal $s < \sfrac{1}{2}$ with the probability of detecting a photon while the signal is smaller than $\sfrac{1}{2}$:
\begin{equation} \label{eq19}
    C_0 \approx \lambda T (1-2 \lambda \tau)(1+\lambda \tau \e^{-2\lambda \tau})
\end{equation}

Incrementing $C_1$ requires detecting a second photons within a time $2\tau$ of the first photon, with an empirical model:
\begin{equation} \label{eq20}
    C_1 \approx C_0 \lambda \tau \e^{2 \lambda \tau}
\end{equation}

Note in Fig.~\ref{fig4} that these models (depicted by the solid lines) describe more accurately the observed behavior of Counter 0 and Counter 1, compared to existing models.

The ratio of the two counters is thus:
\begin{equation} \label{eq21}
    r=\frac{C_1}{C_0}=\lambda \tau \e^{2 \lambda \tau}
\end{equation}
and, including an empirical correction factor of $0.91$, results in:
\begin{equation} \label{eq22}
    \lambda \tau=0.91\frac{\W_{0}(2 r)}{2}
\end{equation}
where $\W_{0}$ is the principal ($0^{th}$) solution of the Lambert W function. Finally, substituting $\lambda \tau$ in Eq.~\ref{eq23} with the solution in Eq.~\ref{eq22} yields the linearization:
\begin{equation} \label{eq23}
    N = \frac{C_0}{(1-2\lambda\tau)(1+\lambda\tau\e^{-2\lambda\tau})}
\end{equation}

\subsection{Novel Empirical Model} \label{sec.3.5}

\begin{figure}[t]
    \includegraphics[width=\columnwidth]{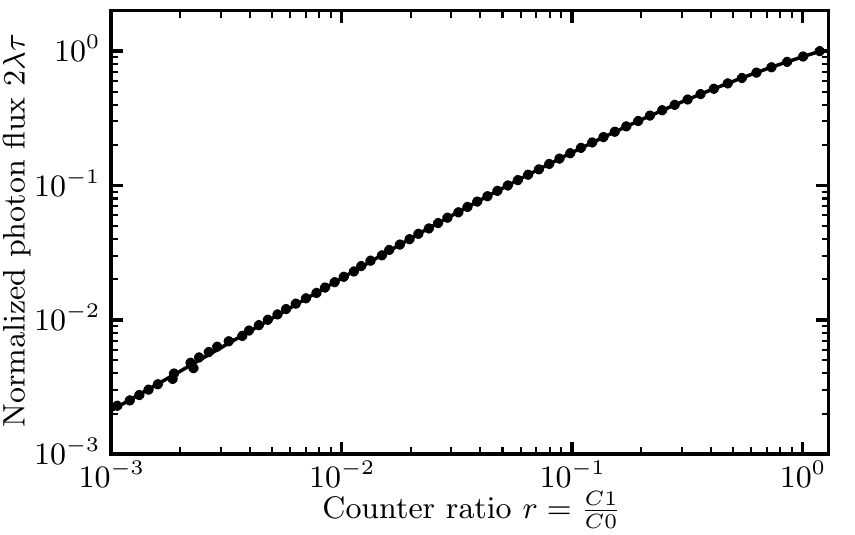}
    \caption{Empirical fitting of the normalized photon flux $2 \lambda \tau$ as a function of counter ratio $r$; black dots indicate results of Monte Carlo simulation and the black line depicts the value of fitting function $f(r)$.}
    \label{fig5}
\end{figure}

\begin{figure}[t]
    \includegraphics[width=\columnwidth]{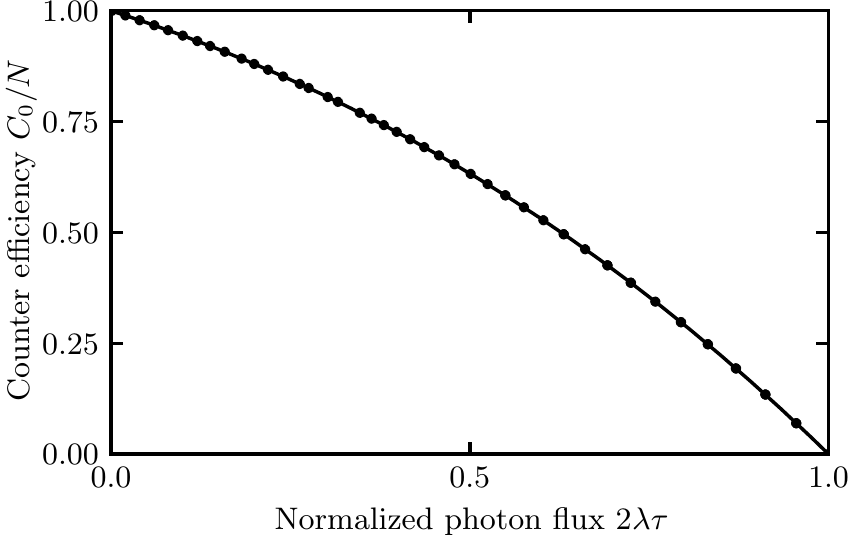}
    \caption{Empirical fitting of the counter $C_0$ efficiency as a function of the normalized photon flux $2 \lambda \tau$; black dots indicate numerical results of Monte Carlo simulation and the black line depicts the value of fitting function $g(\lambda \tau)$.}
    \label{fig6}
\end{figure}

First note that the ratio $r=\frac{C_1}{C_0}$ increases monotonically; we obtain an empirical function:
\begin{equation} \label{eq24}
   \lambda \tau=f(r)=\exp\bigg[{\sum_{i=0}^{3} a_i (\ln{r})^i}\bigg]
\end{equation}
see fitting in Fig.~\ref{fig5} and resulting parameters in Table~\ref{table1}. We can also model the ratio $C_0/N$ as a function of $\lambda \tau$:
\begin{equation} \label{eq25}
    \frac{C_0}{N}=g(\lambda \tau)=\sum_{i=1}^{4} b_i (1-2\lambda\tau)^i
\end{equation}
where $b_4=1-b_1-b_2-b_3$, see fitting in Fig.~\ref{fig6} and resulting parameters in Table~\ref{table2}. From the ratio $r$ we can thus obtain the linearized counting rate:
\begin{equation} \label{eq26}
N = \frac{C_0}{g\big(f(r)\big)}	
\end{equation}
Knowledge or calibration of $\tau$, $T$ and $\lambda$ is not required.

\begin{table}[t]
\caption{Fitting parameters for $f(r)$}  \label{table1}
    \begin{tabular}{rrr}      
        PARAMETER & VALUE & $\sigma$ \\
        $a_0$ & $-0.7908$ & $ 0.0008$ \\
        $a_1$ & $ 0.5500$ & $ 0.0016$ \\
        $a_2$ & $-0.0822$ & $ 0.0008$ \\
        $a_3$ & $-0.0050$ & $ 0.0001$ \\
    \end{tabular}
\end{table}

\begin{table}[t]
\caption{Fitting parameters for $g(\lambda \tau)$}  \label{table2}
    \begin{tabular}{rrr}      
        PARAMETER & VALUE & $\sigma$ \\
        $b_1$ & $ 1.584$ & $ 0.001$ \\
        $b_2$ & $-0.682$ & $ 0.003$ \\
        $b_3$ & $ 0.088$ & $ 0.005$ \\
    \end{tabular}
\end{table}

\subsection{Novel Model for Synchrotron Fill Patterns with Widely Spaced Buckets}
With widely spaced buckets, photons arrive quasi-simultaneously, greatly simplifying the modeling of the counter behavior. Assuming a rate of $\Lambda$ photons/pixel/pulse and long acquisition times with relatively sparse photons in each pulse, we obtain from the Poisson distribution:
\begin{equation} \label{eq27}
	C_0=\sum_{i=1}^{\infty} \frac{\Lambda^i \e^{-\Lambda}}{i!}=N(1-\e^{-\Lambda})
\end{equation}
\begin{equation} \label{eq28}
	C_1=\sum_{i=2}^{\infty} \frac{\Lambda^i \e^{-\Lambda}}{i!}=N[1-(1+\Lambda)\e^{-\Lambda}]
\end{equation}
with ratio
\begin{equation} \label{eq29}
	1-r = 1-\frac{C_1}{C_0} =\frac{\Lambda\e^{-\Lambda}}{1-\e^{-\Lambda}}
\end{equation}
and solution:
\begin{equation} \label{eq30}
	\Lambda=	(r-1)-\W_{-1}((r-1)\e^{r-1})
\end{equation}
where $\W_{-1}$ is the -1\textsuperscript{st} solution of the Lambert W function. This solution can be used to correct the number of pulses in $C_0$:
\begin{equation} \label{eq31}
	N=\frac{C_0}{1-\e^{-\Lambda}}=\frac{C_0}{1-\frac{r-1}{\W_{-1}((r-1)\e^{r-1})}}
\end{equation}

Note that this result is applicable for light sources with (relatively) constant intensity of individual pulses. This is typically not true at free-electron laser sources, where either a more complex gamma distribution should be used instead of the Poisson distribution \cite{blaj2017optimal}, or individual pulses should be read out separately.

\section{Results} \label{sec.4}

In Fig.~\ref{fig7} we show the typical response of counter $C_1$ along with the effect of the various counting rate corrections, including the counting statistics noise (obtained by simulating 100 acquisitions at each point, applying corrections to each result, and using the mean and the standard deviation of each set to estimate the systematic deviations (markers) and the counting noise (error bars), respectively. The ideal response is indicated by the thick solid blue line at $y=1$; models closer to the ideal perform better (i.e., their systematic deviation is smaller).

\subsection{One Counter} \label{sec.4.1}

Linearizing the response of single photon counting pixel detectors with a single counter requires knowledge of the integration time $T$ and dead-time $2 \tau$, and can only be used for $\lambda \ll \frac{1}{4\tau}$. This could be achieved by solving Eq.~\ref{eq4} using the (assumed known) $T$ and $\tau$ parameters, resulting in an estimated number of photons:
\begin{equation}\label{eq32}
	N = -\frac{T}{\tau}	\W_0 \bigg(-\frac{C_0 \tau}{T}\bigg)
\end{equation}

In practice, an empirical calibration is likely to yield superior results, due to the mismatch between the model in Eq.~\ref{eq4} and the actual behavior of $C_0$. The calibration would depend on the parameters $\tau$ and $T$, as well as on the fill pattern of the light source, see discussion in section~\ref{sec.4.3}.

\subsection{Two Counters} \label{sec.4.2}

Using two counters eliminates the need to know or calibrate $E_0$ (photon energy), $2\tau(E_0)$ (time for the analog signal to return from $1$ to $0$) and $T$ (integration time). More accurate models perform well at rates approaching $2\lambda\tau \to \frac{1}{2}$. Using any linearization approach yields superior results compared to no linearization. Even the simple sum of two counters exceeds the performance of the analytical 1 counter model (which also requires knowledge of $E_0$, $\tau(E_0)$ and $T$).

\begin{figure*}[t]
    \includegraphics[width=\textwidth]{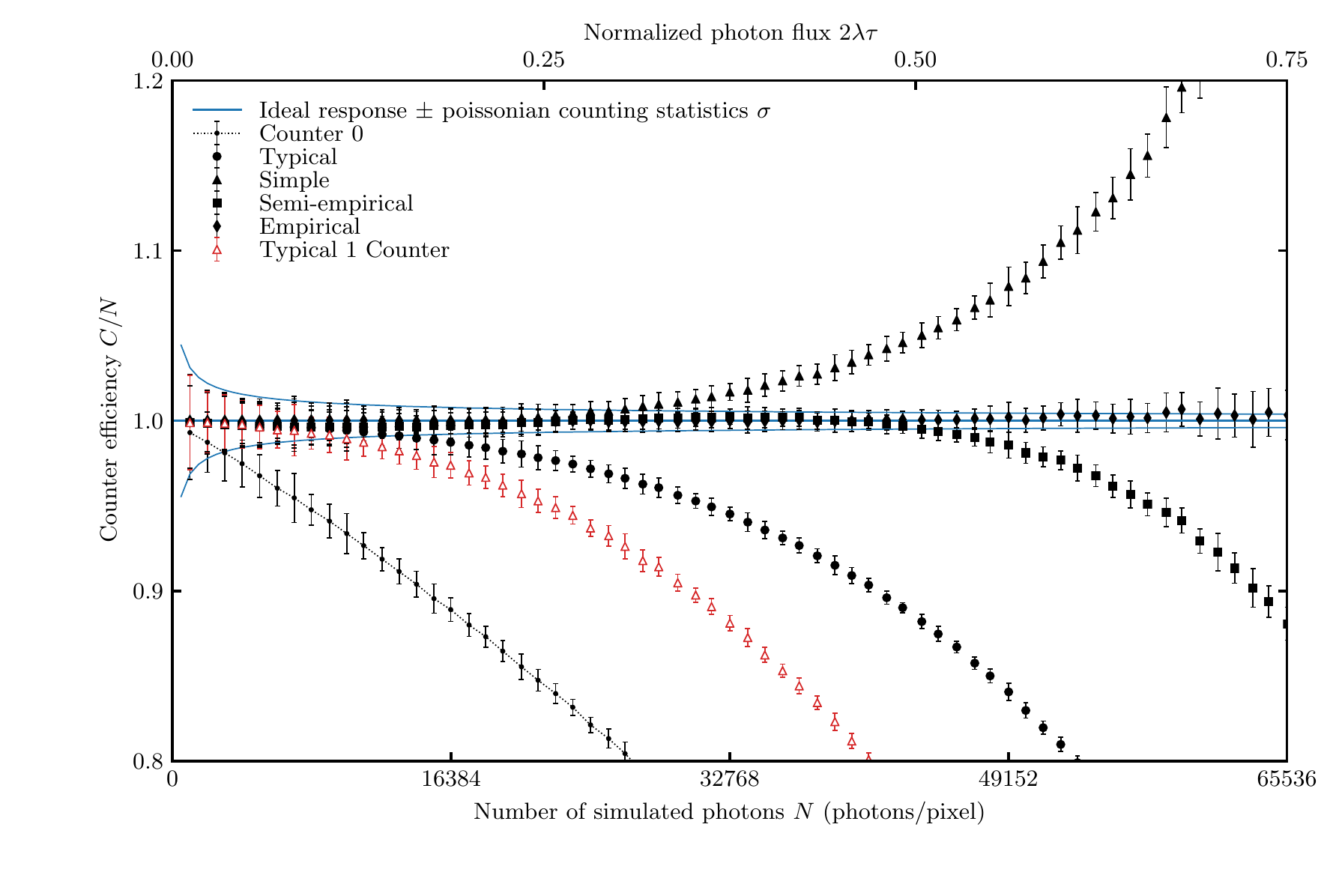}
    \caption{Comparison of counting efficiencies with/without corrections for realistic parameters ($\tau$=100~ns, integration time $T$=20~ms, $\lambda \in$ [$0$, $3.25$~Mcount/pixel/s], pixel counter depth $65536$). The ideal counter efficiency is $1$ over the whole range (thick solid blue line), with the counting statistics limits $\pm \sigma$ indicated by thin solid blue lines. Error bars represent the counting statistics errors after linearization. The uncorrected counter quickly deviates from the ideal (small circles with dotted line). A corrected single counter (open-faced red triangles) requires knowledge of parameters $\tau$ and $T$ and underperforms a simple sum of the two counters (large circles). Various two counter dead time corrections yield increasingly accurate results, extending the linear counting rate by 1 to 2 orders of magnitude (Empirical model, black diamonds) and enabling linear counting rates in excess of $10^7$~counts/pixel/s. }
    \label{fig7}
\end{figure*}

The Simple model (indicated by filled triangles in Fig.~\ref{fig7}) offers an increased linear range with minimal implementation effort and outperforms all previous models; note however the overestimation of the counting rates.

The Semi-empirical and the Empirical models (depicted in Fig.~\ref{fig7} by filled squares and diamonds, respectively) offer the best performance; in particular the empirical model exceeds by far the performance of the other models. Using the Semi-empirical or the Empirical linearization methods with, e.g., the Medipix3RX with a maximum (i.e., nonlinear) counting rate of $4.5$~$10^6$~counts/pixel/second \cite{ballabriga2016review} results in a linear counting rates in excess of $10^7$~counts/pixel/second.

In Table~\ref{table3}, the performance of each linearization model is summarized; the linear range is defined as the range where systematic deviations are limited to the photon counting statistics (indicated by the thin solid blue lines in Fig.~\ref{fig7}). The empirical model yields 1 to 2 orders of magnitude improvement in the linear range compared to a single counter, with an error comparable with the counting statistics for normalized photon rates up to $2\lambda\tau \approx$~$0.6$.

\begin{table*}[t]
\caption{Performance of Photon Counting Linearization Methods}\label{table3}
    \begin{tabular}{llcr}      
        MODEL & EQUATION & SECTION & LINEAR RANGE\\
        \hline
        None 1 Counter & $C_0$ & 2.3 & 3.5\% \\
        Typical 1 Counter & $N(C_0,\tau,T)$ & 4.1 & 11.0\% \\ 
        \hline      
        Typical & $C_0 + C_1$ & 3.1 & 15.1\% \\
       Novel Simple & $C_0/(1-r)$ & 3.2 & 29.7\% \\
       Novel Simple Gain & $C_0/(1-r)^{\frac{2g-\sfrac{3}{2}}{g-\sfrac{1}{2}}}$ & 3.3 & 29.7\% \\
       Novel Semi-empirical & $C_0/[(1-2 x)(1+x\e^{-2 x})]$& 3.4 & 50.0\% \\
       Novel Empirical & $C_0/g(f(r))$ & 3.5 & 64.5\% \\
        \hline
    \end{tabular}\\
    where $r=C_1/C_0$ and $x=0.91 W(2 r)/2$, with $\W$ the Lambert W function.
\end{table*}

\subsection{Synchrotron Fill Patterns} \label{sec.4.3}

Synchrotron fill patterns influence the behavior of the counters. For regular fill patterns and closely spaced buckets (e.g., typical gaps $\delta t \approx$ 2~ns at synchrotrons, much smaller than dead-times $\tau \approx$~100~ns ($\delta t \ll \tau$), the formulas above remains valid.

With larger gaps approaching $\tau$ ($\delta t < \tau/2$), the equations describing the behavior of individual counters have to be corrected: let $f$ be the fraction of time with beam and $1-f$ the fraction of time with gaps; then we can substitute $\lambda$ with $\lambda/f$ and $T$ with $f T$. Note that in detectors with two thresholds $\tau$, $T$ and $f$ drop out of the linearization equations \ref{eq3}, \ref{eq7}, \ref{eq14}, \ref{eq23} and \ref{eq26}, leading to exactly the same result.

For large intervals between buckets, $\delta t \gg \tau$, the dead-time is irrelevant and the pile-up is reduced to the elegant solution in Eq.~\ref{eq31}.

Finally, for other fill patterns (e.g., $\delta t \approx \tau$), the Poisson distribution (or gamma distribution, in case of free-electron laser beams) in each bucket needs to be estimated. Alternatively, the correspondence between the ratio $C_1/C_0$ and the ratio $N/C_0$ can be calibrated empirically as a function of beam intensity as shown in section~\ref{sec.3.5}, yielding a calibration curve specific to the actual fill pattern.

\section{Conclusions}

In the last decade, a revolution in hybrid pixel detectors led to the wide availability of photon counting pixel detectors that enabled rapid advances in photon science and applications of high dynamic range imaging.

In the near future, significant brilliance upgrades are expected for most storage ring X-ray sources \cite{chenevier2018esrf}. High repetition rate free-electron lasers might also eventually operate in a (quasi-)continuous wave mode \cite{marcus2017lcls,brinkmann2014prospects}. This will further increase the pressure to linearize the saturation response of photon counting pixel detectors, or possibly replace them with fast integrating detectors.

One existing approach to linearize the photon counting response is to assume a simple paralyzable model response with a fixed dead-time $\tau$; however, the typical photon counting detector with multiple thresholds has a constant current discharge of the feedback capacitor, resulting in very different statistics. Another existing approach to linearize the photon counting response is to combine a single counter with a time-over-threshold counter. This approach increases counting linearity, however, its range is limited by the depth of the time-over-threshold counter and the signal-to-noise performance is limited by the higher noise in time-over-threshold compared to photon counting.

In this paper we introduced new, more accurate statistical descriptions of the dead-time and photon pile-up in photon counting detectors with multiple thresholds, describing the counting behavior of typical photon counting detectors with two independent counters (corresponding to two different thresholds). Subsequently, we derived three nonlinearity correction formulas, ranging from simple to complex, with the important advantage that they significantly extend the region with linear photon counting compared with previous approaches.

Additionally, in detectors with at least two thresholds and counters, these linearization methods are very robust and do not require a calibration, or knowledge of the individual system parameters (integration time, dead-time, etc.), thus enabling simple implementation, or easy re-processing of previous experimental data.

Finally, we described the influence of synchrotron fill patterns on the photon counting linearization methods, showing that they can be applied in a majority of cases (closely spaced buckets, approaching typical detector dead times of $\approx$ 100~ns). We developed a new model for widely spread buckets. Finally, for bucket spacing similar to the detector dead time, we described an empirical calibration method.

These novel methods enable extending by $1$ to $2$ orders of magnitude the range of linear counting for photon counting detectors with a constant current discharge.


     
\begin{acknowledgements}
Work supported by the U.S. Department of Energy, Office of Science, Office of Basic Energy Sciences under contract number {DE-AC02-76SF00515}.

The author would like to thank R.~Ballabriga and L.~Tlustos at CERN for many interesting discussions on photon counting and pixel detectors.
\end{acknowledgements}


%

\end{document}